\documentclass{aa}

\usepackage{graphicx}
\usepackage{txfonts}

\newcommand{\Mpc}{$h^{-1}$\thinspace Mpc}
\newcommand{\etal}{{\rm et al.~}}

\def\apjl{ApJL}
\def\apjs{ApJS}

\def\aap{A\&A}

\begin{document}

\title{Luminous superclusters: remnants from inflation}

\author{ J. Einasto\inst{1} \and M. Einasto\inst{1} \and
 E. Saar\inst{1} \and  E. Tago\inst{1} \and L. J. Liivam\"agi\inst{1} \and
 M. J\~oeveer\inst{1} \and I. Suhhonenko\inst{1} \and
 G. H\"utsi\inst{1} \and J. Jaaniste\inst{2} \and P.
 Hein\"am\"aki\inst{3} \and V. M\"uller\inst{4} \and A. Knebe\inst{4}
 \and D. Tucker\inst{5}}

\institute{Tartu Observatory, EE-61602 T\~oravere, Estonia
\and 
Estonian University of Life Sciences
\and 
Tuorla Observatory, V\"ais\"al\"antie 20, Piikki\"o, Finland 
\and
Astrophysical Institute Potsdam, An der Sternwarte 16,
D-14482 Potsdam, Germany
\and
 Fermi National Accelerator Laboratory, MS 127, PO Box 500, Batavia,
IL 60510, USA
}

\date{ Received 2006; accepted} 

\authorrunning{J. Einasto et al.}

\titlerunning{Luminous superclusters}

\offprints{J. Einasto }

\abstract { We derive the luminosity and multiplicity functions of
  superclusters compiled for the 2dF Galaxy Redshift Survey, the Sloan Digital
  Sky Survey (Data Release 4), and for three samples of simulated superclusters.
  We find for all supercluster samples Density Field (DF) clusters, which
  represent high-density peaks of the class of Abell clusters, and use median
  luminosities/masses of richness class 1 DF-clusters to calculate relative
  luminosity/mass functions. We show that the fraction of very luminous
  (massive) superclusters in real samples is more than tenfolds greater than
  in simulated samples. Superclusters are generated by large-scale density
  perturbations which evolve very slowly. The absence of very luminous
  superclusters in simulations can be explained either by non-proper treatment
  of large-scale perturbations, or by some yet unknown processes in the very
  early Universe.


}

\maketitle

\section{Introduction}

Superclusters are the largest density enhancements in the Universe of common
origin. Superclusters evolve slowly and contain information from the very
early Universe.  The investigation of large systems of galaxies was pioneered
by the study of the {\em Local Supercluster} by de Vaucouleurs (\cite{deV53}).
Another approach to define superclusters was initiated by Abell (\cite{abell},
\cite{abell61}), who considered them as {\em ``clusters of clusters''}.  Until
recently, superclusters have been found mostly on the basis of catalogues of
rich clusters of galaxies by Abell (\cite{abell}) and Abell et al.
(\cite{aco}).  Abell supercluster catalogues have been complied by Zucca \etal
(\cite{z93}), Einasto \etal (\cite{e1994}, \cite{e1997}, \cite{e2001}) and
Kalinkov \& Kuneva (\cite{kk95}).

Actually superclusters consist of galaxy systems of different richness: single
galaxies, galaxy groups and clusters, aligned to chains (filaments).  This has
already been realized before by J\~oeveer, Einasto \& Tago (\cite{jet78}),
Gregory \& Thompson (\cite{gt78}), Zeldovich, Einasto \& Shandarin
(\cite{zes82}), and has been confirmed by recent studies of superclusters
using new deep galaxy surveys, such as the Las Campanas Galaxy Redshift
Survey, the 2 degree Field Galaxy Redshift Survey (2dFGRS, Colless et al.
\cite{col01}, \cite{col03}) and the Sloan Digital Sky Survey Data Release 4
(SDSS DR4, Adelman-McCarthy et al. \cite{dr4}).  New galaxy redshift surveys
are almost complete in a fixed apparent magnitude interval.  This allows to
estimate total luminosities of superclusters using weights, inversely
proportional to the number of galaxies in the observational window of apparent
magnitudes.  This possibility has been used in recent supercluster studies by
Basilakos (\cite {bas03}), Basilakos et al.  (\cite{bpr01}), Erdogdu et al.
(\cite{erd04}), Porter and Raychaudhury (\cite{pr05}), and Einasto et al.
(\cite{e03a}, \cite{e03b}, \cite{ein05}, \cite{e06a}, hereafter Paper I).

\begin{figure*}[ht]
\centering
\resizebox{.48\textwidth}{!}{\includegraphics*{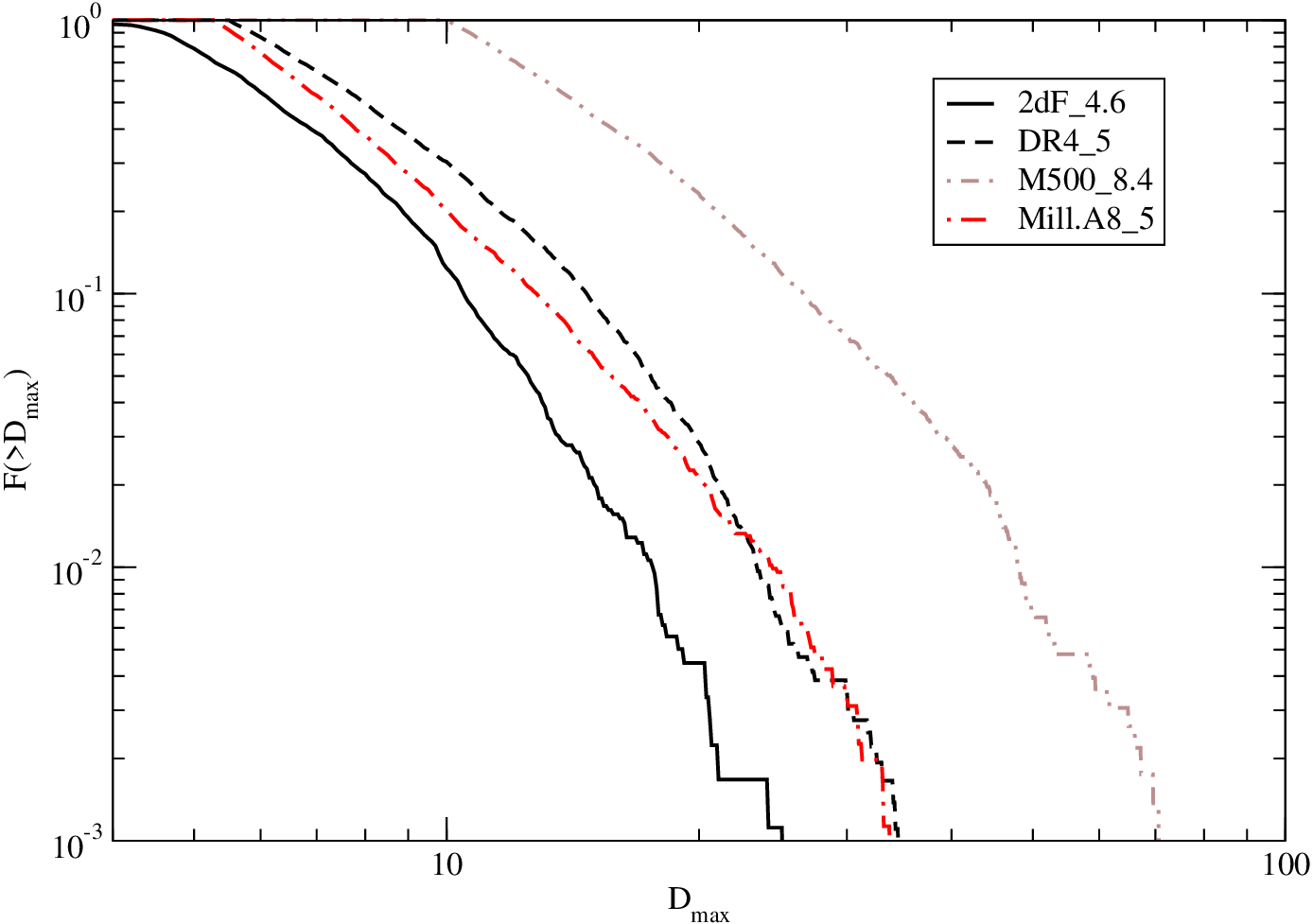}}
\resizebox{.48\textwidth}{!}{\includegraphics*{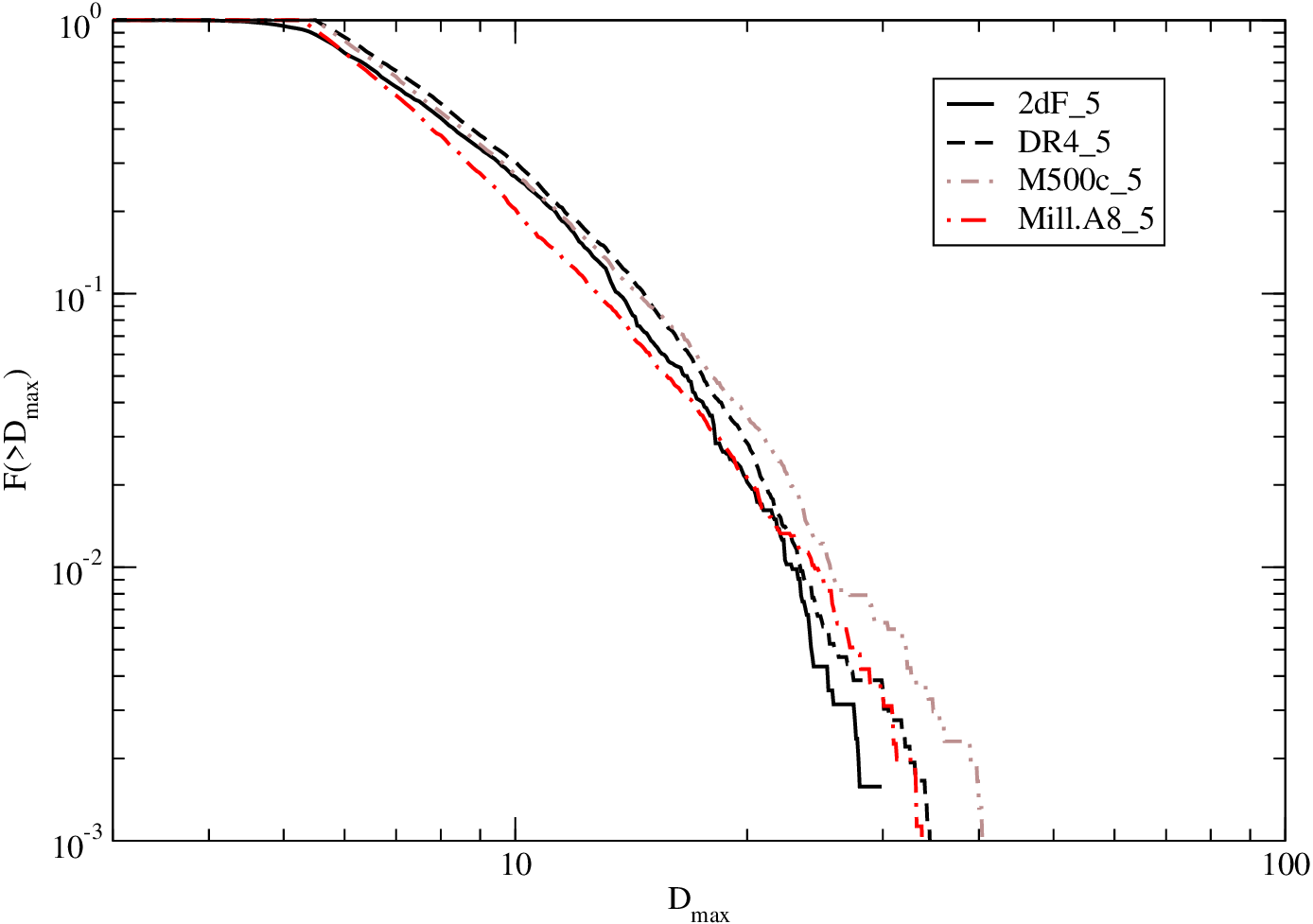}}
\caption{The cumulative distribution of peak densities of DF-clusters. Left
  panel shows densities uncorrected for the relative bias, right panel
  densities corrected for differences in the relative bias.
\label{fig:1}}
\end{figure*}

In Paper I we compiled a catalog of superclusters using the 2dFGRS. A similar
catalogue on the basis of SDSS Data Release 4 (DR4) is in preparation (Einasto
et al.  \cite {e06c}, hereafter E06c).  The properties of 2dFGRS superclusters
were analyzed by Einasto et al.  (\cite{e06b}, Paper II).  The parameters of
2dFGRS superclusters were compared with properties of model superclusters
based on the Millennium Simulation of the evolution of the Universe by
Springel et al.  (\cite{springel05}).  This comparison of real superclusters
with simulated ones shows that geometric properties of simulated superclusters
agree very well with similar properties of real superclusters. This
demonstrates that the ability to simulate processes which lead to the
formation of superclusters has reached an advanced stage.  However, one
property of model superclusters is in conflict with reality: real samples have
many more very luminous superclusters than model samples.  The presence of
very massive superclusters in our neighborhood is well known, examples are the
Shapley and Horologium-Reticulum Superclusters (see Fleenor et al.
\cite{fleenor05}, Proust et al.  \cite{proust06}, Nichol et al.
\cite{nichol06} and Ragone et al.  \cite{ragone06} and references therein).
However, the number of such extremely massive superclusters was too small to
make definite conclusions on the phenomenon.

The goal of this Letter is to determine the luminosity and
multiplicity functions of 2dFGRS and SDSS DR4 superclusters and to
compare these functions with similar functions of simulated
superclusters.  These observational samples of superclusters are the
largest available today.  To verify the robustness of the results
obtained with the Millennium simulation data we also use superclusters
derived from a cosmological simulation of the same volume but with a
lower mass resolution.  This large collection of real and model
superclusters allows us to make definite conclusions on the statistics
of luminous superclusters.  Supercluster catalogues of the 2dFGRS, as
well as fits files of luminosity density fields of 2dFGRS and
Millennium Simulation are available electronically at the web-site
\texttt{http://www.aai.ee/$\sim$maret/2dfscl.html}.

\section{Data}

In this analysis we used galaxy and group samples of the 2dFGRS (Tago
et al. \cite{tago06a}, Paper I) and SDSS DR4 (Tago et
al. \cite{tago06b}, E06c), and three simulated samples.  To compile
supercluster catalogues we used the density field method which allows
to correct the field to take into account the presence of faint
galaxies outside the observational window, for details see Paper I.
The richness of a supercluster can be characterised by its total
luminosity, and by the number of rich galaxy clusters in it, i.e.  the
multiplicity function of the supercluster.  In Papers I and II we
derived both these characteristics.  The sample of Abell clusters
covers only our close neighbourhood, in more distant superclusters
they are found only in exceptional cases. To have a richness parameter
we used instead of Abell clusters high-density peaks of the density
field, which we call DF-clusters.  The spatial density of DF-clusters
is approximately close to the spatial density of Abell clusters, thus their
utilization as richness indicator yields results comparable with
previous supercluster studies based on Abell clusters.

For comparison to the observational data sets we used the galaxy catalogues
based upon the Millennium Simulation, for details see Springel et al.
(\cite{springel05}), Gao et al.  (\cite{gao05}) and Croton et al.
(\cite{croton06}).  In addition, we also performed a cosmological dark matter
simulation of a computational volume of side-length 500~\Mpc\ using $256^3$
particles (model M500). We chose the concordance cosmology with the parameters
$\Omega_m = 0.27$, $\Omega_{\Lambda} = 0.73$, $\sigma_8 =0.84$. The simulation
was carried out with the open source Multi Level Adaptive Particle Mesh
(MLAPM) code by Knebe \etal\ (\cite{knebe01}) and DM-halos have been found by
the conventional Friends-of-Friends (FoF) procedure.  We denote simulated
samples as Mill.A8 and M500.  The density fields of these models were calculated
using all simulation galaxies/particles.  In addition, we used the model
Mill.F8,  where galaxies were chosen using similar selection criteria as
in the 2dFGRS sample (see Paper I).

To get comparable results for the luminosity and multiplicity
functions of different samples all data and data reduction procedures
must be as similar as possible.  First of all, density fields must
have identical threshold bias levels.  We use in our study relative
densities expressed in units of the mean density of the particular
sample.  To check the relative bias levels we found for all samples
the threshold density, which yields non-percolating superclusters of
maximal diameter approximately 100 - 120 \Mpc, and selected
DF-clusters using a threshold density about 10 \% higher than used in
the selection of superclusters.

{\scriptsize
\begin{table*}[ht]
\caption[]{Data on supercluster samples}
\label{tbl}
      \[
\begin{tabular}{l|rr|rrrccc|rrrccc}
\hline
\noalign{\smallskip}
Sample  &\multicolumn{2}{c|}{}&
            \multicolumn{6}{c|}{$D_0 = 5.0$}&
            \multicolumn{6}{c}{$D_0 = 6.0$} \\  
& $V$ & $N_{gal}$& $N_{cl}$ &$N_{scl}$&$N_{1}$&$L_0$ & $n_{cl}$& $n_{scl}$&
$N_{cl}$ &$N_{scl}$&$N_{1}$&$L_0$ & $n_{cl}$& $n_{scl}$\\ 
            \noalign{\smallskip}
            \hline
            \noalign{\smallskip}
2dFc    & 30.3&  184395& 2555&  567&  279& 6.86e+11& 84& 19& 1664&  544& 265&
8.20e+11& 55& 18 \\
SDSS    & 43.9&  368002& 3621& 1012&  517& 7.09e+11& 82& 23& 2364&  911& 483&
8.11e+11 & 54& 21\\
Mill.A8    &125.0& 8964936& 4914& 2259& 1292& 1.64e+12& 39& 18& 2878& 1733&1025&
2.10e+12 & 23& 14\\
Mill.F8   &125.0&2094187 & 3020 & 1299& 762& 3.08e+12& 24& 10& 1687 & 1068&
752& 3.03e+12& 14& 8 \\
M500c   &125.0& 9785827& 3032& 1860& 1281& 4.75e+14& 24& 15& 1880& 1440&1037&
5.81e+14 & 15& 12\\ 
            \noalign{\smallskip}
            \hline
\end{tabular}
      \]
\end{table*}
}

The comparison of peak densities of DF-clusters in differing samples
used in the present study shows considerable disparities amongst the
samples, see Fig.~\ref{fig:1}.  These variations are due to
differences in mean densities used in the calculation of relative
densities, i.e. differences in the threshold bias factor (for a detailed
discussion of this phenomenon see Einasto et al.  \cite{e1999}).
Fig.~\ref{fig:1} shows that peak density distributions of DF-clusters of the
SDSS and Millennium simulation samples are practically identical, thus we have
used these samples as a standard.  To bring the threshold biases of the other
samples to a comparative level we divided the density fields of the 2dFGRS
samples by 0.739, and the model M500 density field by 2.025 (these values were
found by comparing peak density distributions shown in the left panel of
Fig.~\ref{fig:1}).  The distribution of peak densities of the corrected
density fields is shown in the right panel of Fig.~\ref{fig:1}.  We see that
there are practically no systematic differences between various samples.
These corrected density fields were used to compile the supercluster
catalogues used throughout this study.

To obtain comparable results we used for all samples identical
procedures in the preparation of the data.  In all cases superclusters
were found using a luminosity (or mass) density field smoothed with an
Epanechnikov kernel of radius 8~\Mpc.  For observational samples
densities were calculated using weights of galaxies which take into
account galaxies and galaxy groups too faint to fall into the
observational window of absolute magnitudes at the distance of the
galaxy.  This is the conventional approach for obtaining the
luminosity density field (Basilakos et al.  \cite{bpr01}, Paper I).
The density field was found for a cell size of 1~\Mpc, which allows to
investigate in detail the internal structure of superclusters (see
Paper II).  Superclusters were defined as connected non-percolating
systems with densities above a certain threshold density.  This
threshold density is similar to the linking length used in the
Friends-of-Friends (FoF) method to find systems of galaxies (or
particles in simulations).

Supercluster catalogues have been compiled for two values of the
threshold density, i.e. $D_0=5.0$ and 6.0 (in units of the corrected
mean density). In both cases a lower limit of the supercluster volume
of 100~(\Mpc)$^3$ was applied. The main characteristics of these
catalogues have been summarized in Table~\ref{tbl}: the name of the
sample (c indicates the use of corrected density fields as described
above), $V$ is the volume (in million cubic \Mpc), $N_{gal}$ is the
number of galaxies (or particles in case of the model M500c) used in
the determination of the density field and the luminosity/mass of
superclusters.  Next we give for both threshold densities the number
of DF-clusters in the sample, $N_{cl}$ (to derive DF-clusters we used
threshold densities 5.5 and 6.5 for the supercluster samples 5.0 and
6.0, respectively). $N_{scl}$ and $N_1$ are the total number of
superclusters, and the number of superclusters of multiplicity 1
(i.e. containing just one DF-cluster), respectively. $L_0$ is the mean
luminosity (or mass for M500c) of superclusters of multiplicity 1,
expressed in Solar units.  These luminosities/masses were used in the
calibration of relative luminosity functions, see the next section for
details.  Finally, $n_{cl}$ and $n_{scl}$ are the mean spatial
densities of DF-clusters and superclusters: the number of objects per
million cubic \Mpc.  
The threshold density 6.0 for the corrected density field
of the 2dFGRS sample corresponds to a threshold density of 4.4 in the
uncorrected density field,  very close to the value 4.6
used in Paper I to find 2dFGRS superclusters. Thus the number of
superclusters and their parameters is very close to the respective
data used in Paper I.

\begin{figure*}[ht]
\centering
\resizebox{.48\textwidth}{!}{\includegraphics*{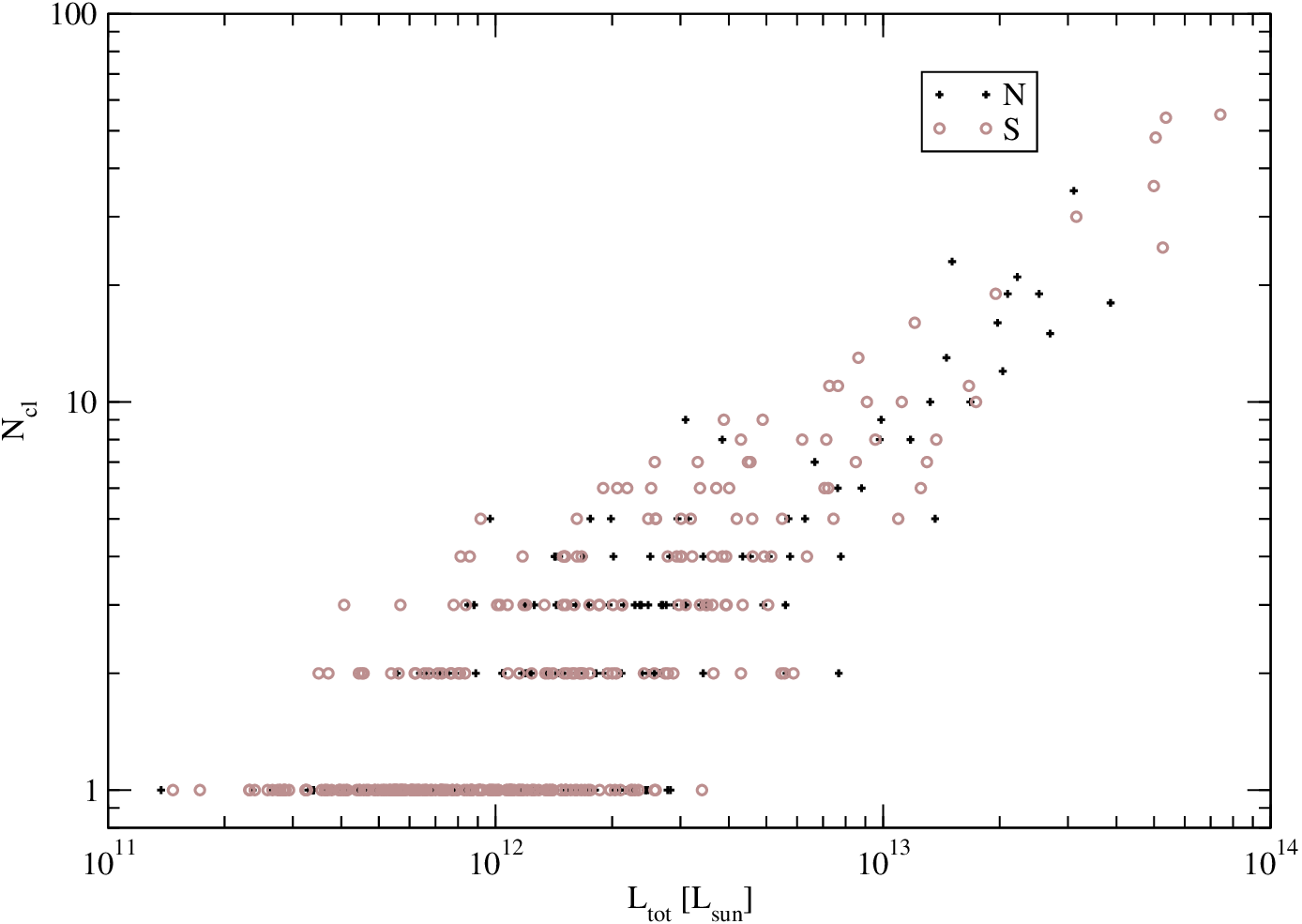}}
\resizebox{.48\textwidth}{!}{\includegraphics*{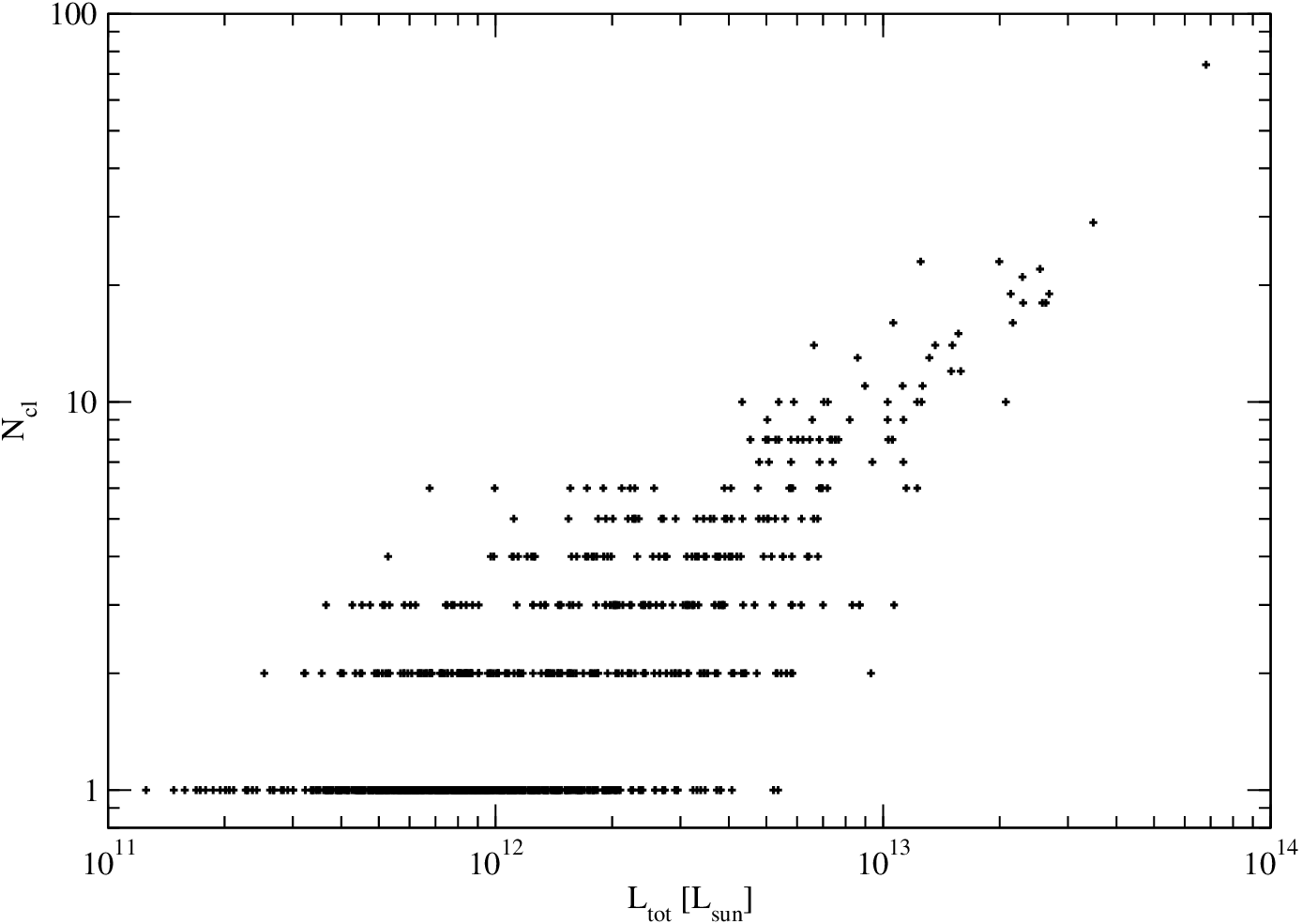}}
\\
\resizebox{.48\textwidth}{!}{\includegraphics*{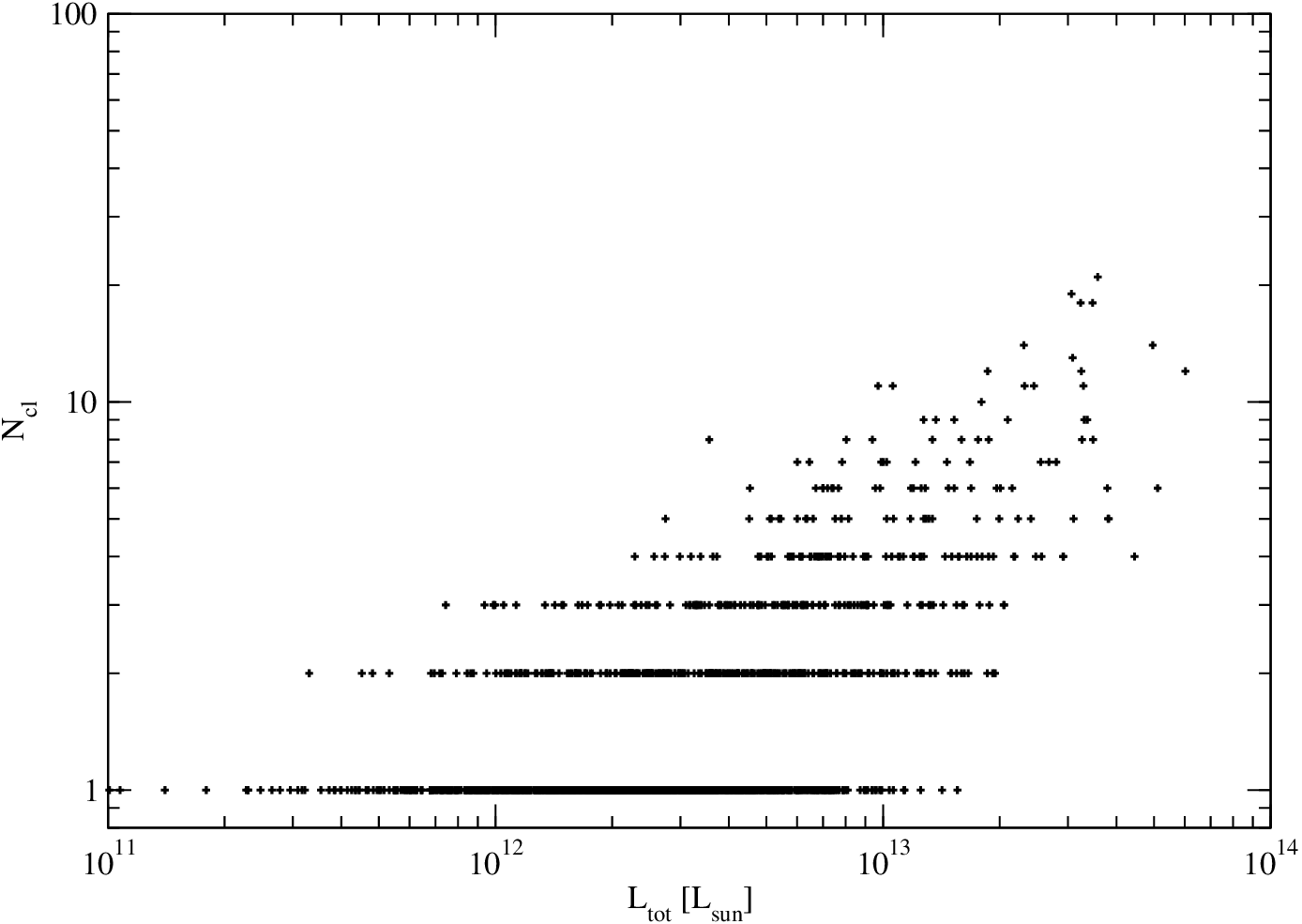}}
\resizebox{.48\textwidth}{!}{\includegraphics*{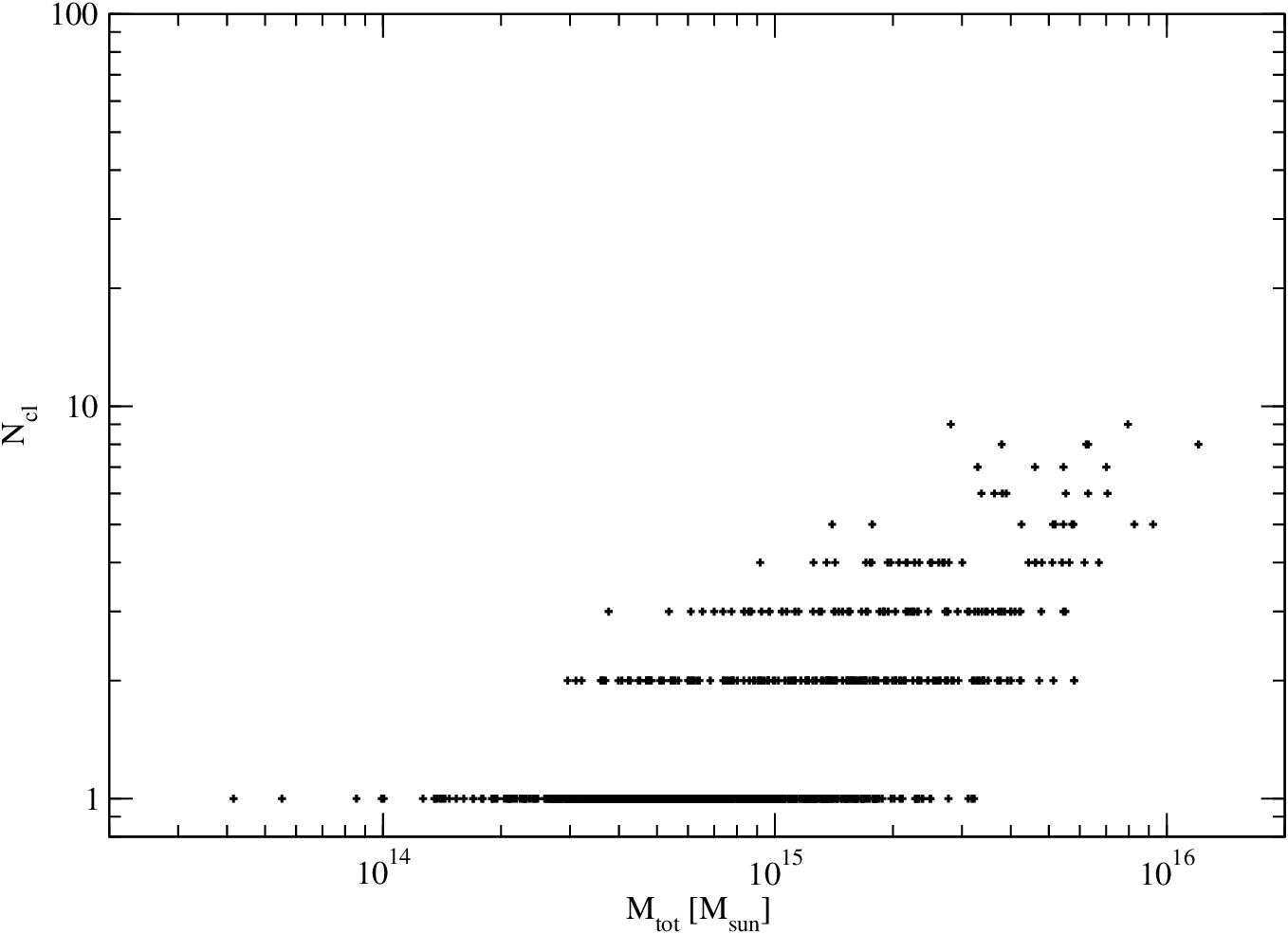}}
\caption{The relationship between the multiplicity and the total luminosity
  (mass) of real and simulated superclusters, calculated for threshold density
  6.0.  The samples are: 2dFGRS (top left), SDSS DR4 (top right), Millennium
  Simulation Mill.A8 (bottom left), the DM simulation M500 (bottom right).
\label{fig:2}}
\end{figure*}

\begin{figure*}[ht]
\centering
\resizebox{0.48\textwidth}{!}{\includegraphics*{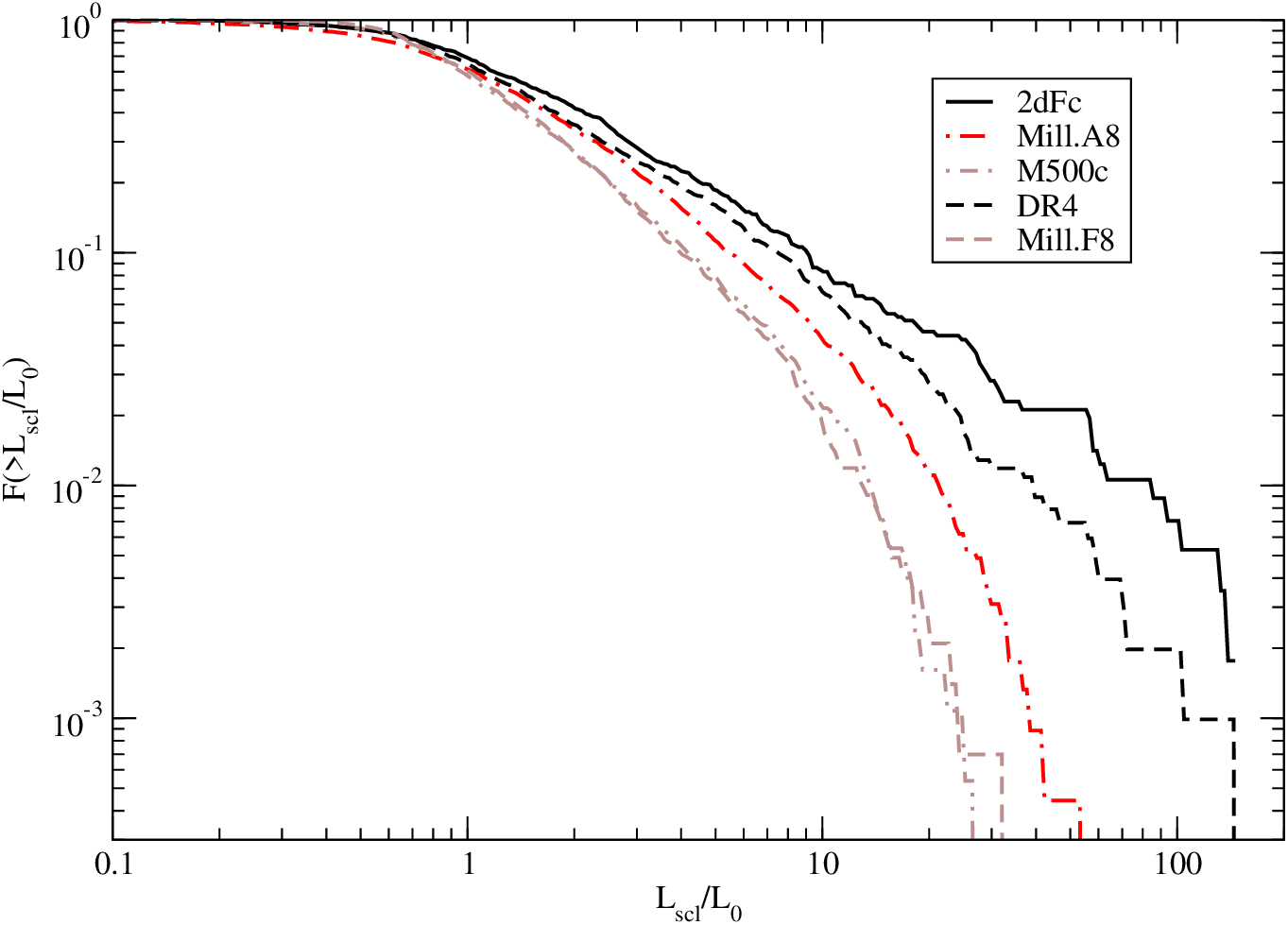}}
\resizebox{0.48\textwidth}{!}{\includegraphics*{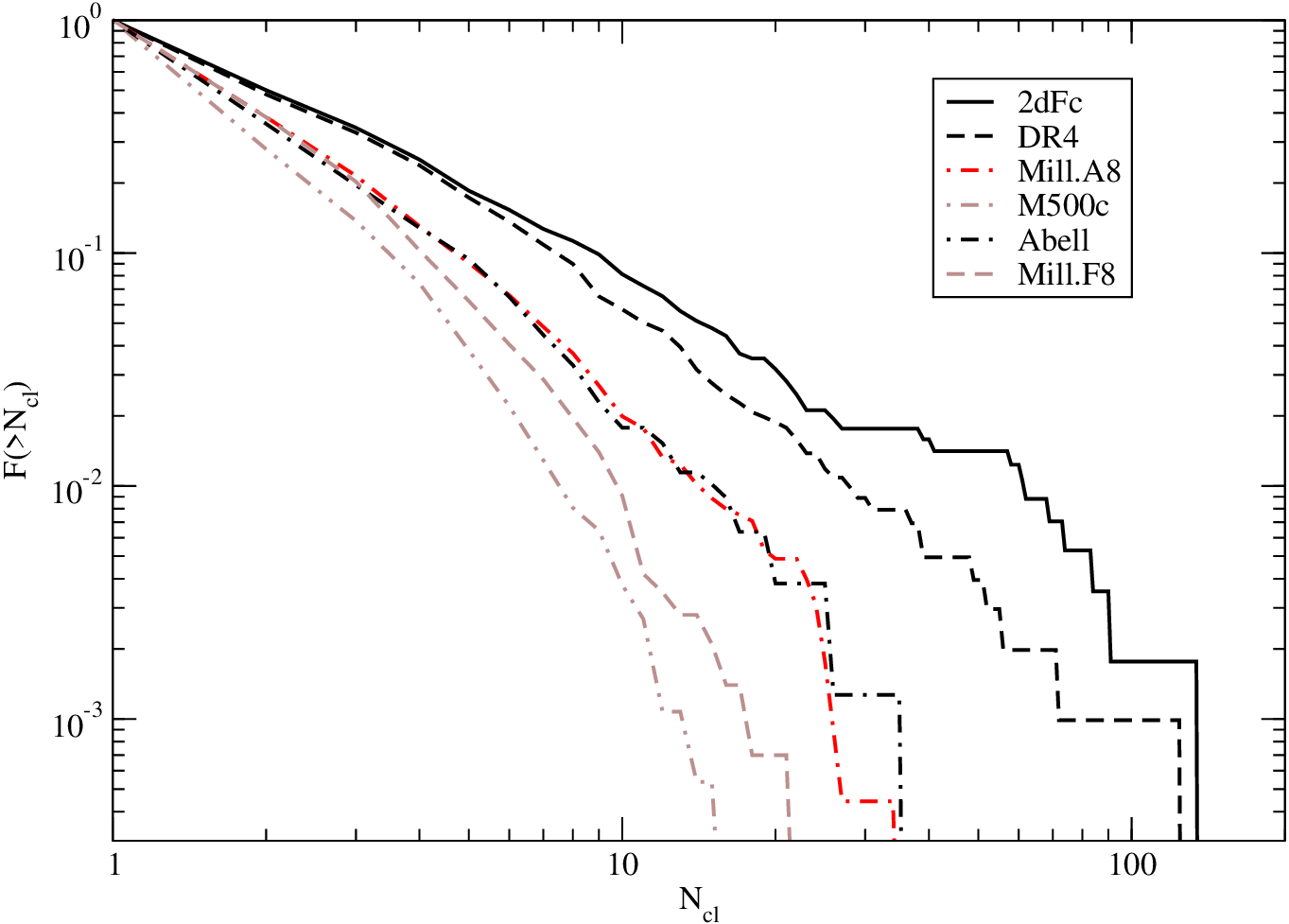}}\\
\resizebox{0.48\textwidth}{!}{\includegraphics*{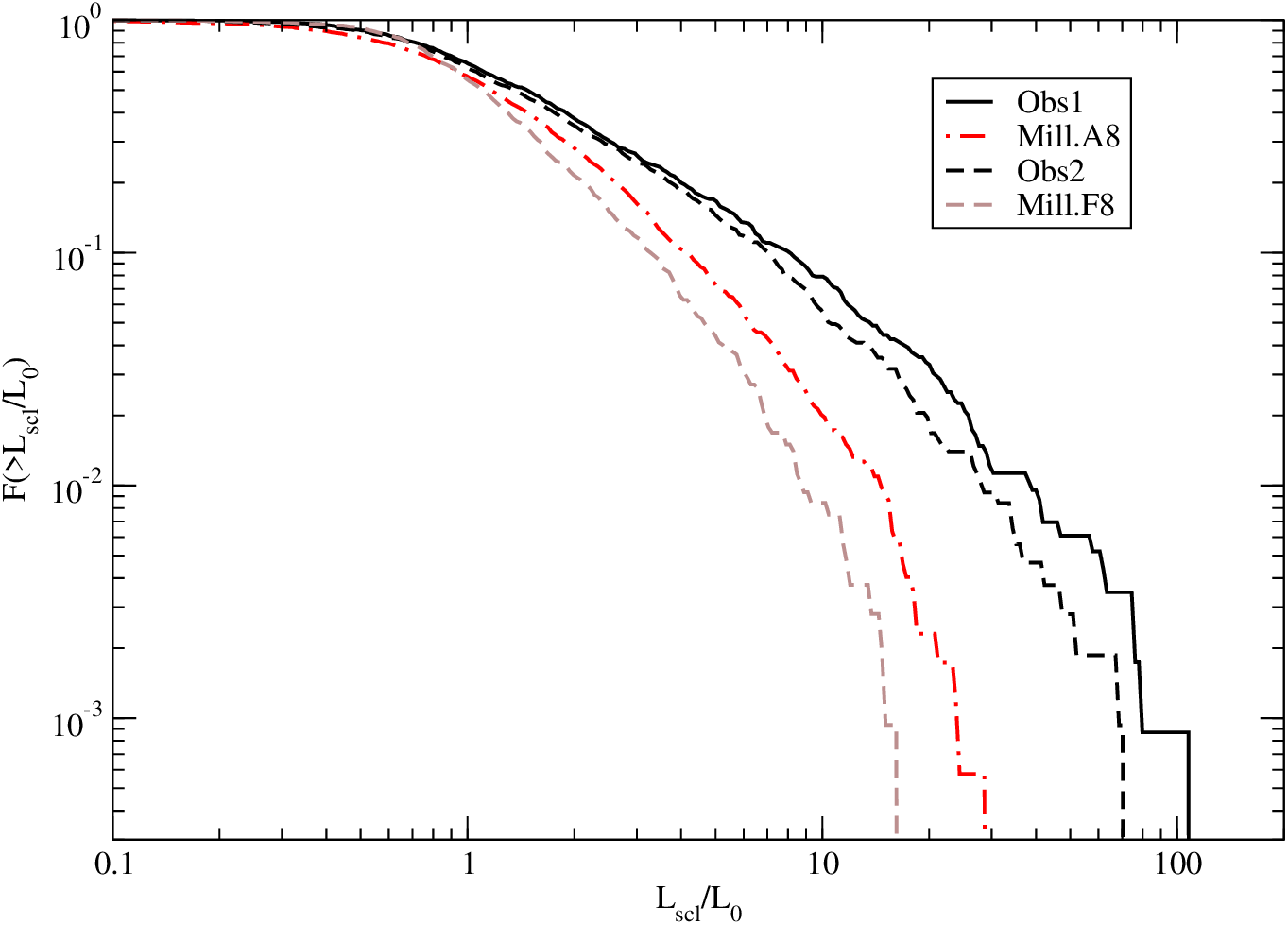}}
\resizebox{0.48\textwidth}{!}{\includegraphics*{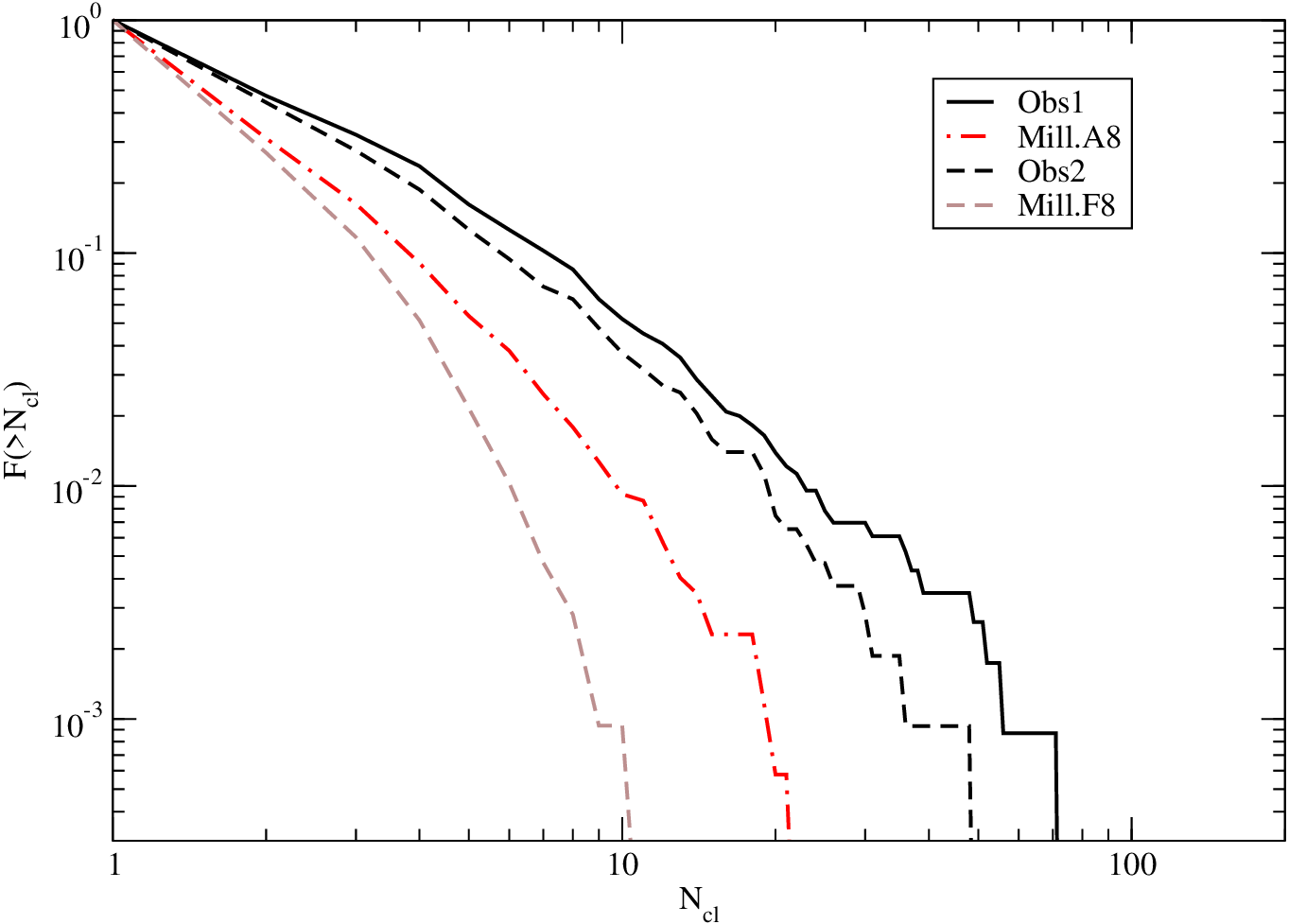}}
\caption{Upper panels show  the relative integrated luminosity/mass functions
  (left), and the multiplicity functions (right) of all real and simulated
  superclusters, found using threshold density 5.0.  Luminosities/masses are
  expressed in units of mean luminosities/masses of superclusters of
  multiplicity 1, $L_0$, given in Table~\ref{tbl}.  Different lines mark
  supercluster samples: the 2dFGRS, the Sloan Digital Sky Survey DR4,
  the Millennium simulation full sample Mill.A8 and selected sample Mill.F8,
  the M500 sample and the sample of Abell superclusters.  Lower panels show
  the comparison of the combined observational samples Obs1 and Obs2 with the 
  Millennium simulation samples Mill.A8 and Mill.F8, using threshold density 6.0.}
\label{fig:3}
\end{figure*}

\section{The luminosity and multiplicity functions of superclusters}

We shall use in this paper two independent parameters to characterise
quantitatively the richness of a superclusters: the multiplicity and the total
luminosity (or mass).  We define the multiplicity of a supercluster by the
number of DF-clusters in it.  DF-clusters are high-density
peaks of the density field, smoothed on a scale of 8~\Mpc.  As seen from
Table~\ref{tbl}, for threshold density $D_0 = 6.0$ the spatial density of
DF-clusters in our samples is about twice the spatial density of
Abell clusters,  25 per million cubic \Mpc\ (Einasto et al. \cite{e1997});
for a threshold density of 5.0 the density is somewhat higher.  The other
integral parameter of a supercluster is its total luminosity or mass,
determined by summing luminosities/masses of all galaxies and groups of
galaxies (DM-particles) inside the threshold iso-density contour which was
used in the definition of superclusters.  The relationship between the
multiplicity and total luminosity (or mass for M500) is presented in
Fig.~\ref{fig:2}.  We see that luminosities have a rather large spread for
superclusters of given multiplicity, the lower the multiplicity the larger the
spread. The other feature, seen in Fig.~\ref{fig:2}, is the shift in the mean
luminosities of superclusters for different samples. Partly this may
be caused by the use of different color systems in various samples.

We are interested in the relative fraction of rich and very rich superclusters
in respect to the number of poor superclusters.  To avoid complications due to
the use of different color systems and masses in case of the model M500, we
define {\em relative} luminosities (masses) as the luminosity (mass) in terms
of the mean luminosity (mass) of poor superclusters, i.e. superclusters that
only contain one DF-cluster and hence are classified as richness class 1.  The
distribution of luminosities is approximately symmetrical on a logarithmic
scale (see Fig.~\ref{fig:2}). We therefore used the logarithm of the
luminosity to derive the mean value.  This value, $L_0$, is also listed for
all samples in Table~\ref{tbl}, in units of the Solar luminosity (mass).

In Fig.~\ref{fig:3} we now show the relative luminosity (mass) functions (left
panels) alongside the multiplicity functions (right panels) for the
observational and model samples.  The spatial density of superclusters is
expressed in terms of the total number of superclusters in the respective
sample to avoid small differences due to the mean number density of
superclusters in different samples. In upper panels these functions are shown
for all observational and model samples using threshold density 5.0, in lower
panels only for the combined observational samples Obs1 and Obs2 (see below),
and model samples Mill.A8 and Mill.F8, using threshold density 6.0.  The
comparison of data obtained with threshold densities 5.0 and 6.0 shows that in
the first case some superclusters of the real sample are actually percolating
(the maximal diameter of the largest superclusters is almost 200~\Mpc).  For
this reason we have used in the final analysis supercluster samples found with
threshold density 6.0.

Upper panels of Fig.~\ref{fig:3} show that there are small differences between
the observational samples 2dFGRS and SDSS: the 2dFGRS sample contains
relatively more rich and very rich superclusters. These differences may be due
to the differing depth of these samples: the limiting magnitude of the 2dFGRS
sample is about 19.35, whereas that of the SDSS sample is 17.7.  Clusters form
superclusters via intermediate density galaxy bridges.  The 2dFGRS sample
contains more faint galaxy bridges between high-density regions which
facilitates the formation of more luminous superclusters.  A similar
difference is observed between the two model samples.  The Mill.A8 sample has
a very high spatial resolution and contains numerous faint galaxies which also
acts in favour of joining nearby high-density regions via galaxy bridges.

To check this explanation of the difference between samples of various depth
we used the simulated 2dFGRS sample Mill.F8, containing only galaxies brighter
than $19.35$, and calculated for this sample luminosity and multiplicity
functions as for other superclusters samples, using threshold densities 5.0
and 6.0.
These versions are also shown in Fig.~\ref{fig:3}.  We see that in this case the
fraction of luminous superclusters in lower than for the full sample Mill.A8.
This test shows that faint galaxy bridges between clusters play indeed an
important role in the formation of superclusters.  To reduce the SDSS DR4
supercluster sample to the same level of bridge strength as the 2dFGRS sample,
a lower threshold density must be used.  Trial calculations showed that a
combined observational sample of superclusters can be formed using the 2dFGRS
sample with threshold density 6.0, and the SDSS DR4 sample with a threshold
density 5.0.  To avoid the inclusion to the combined observational sample data
of lower accuracy we used only superclusters of the main sample with distances
up to 520~\Mpc.  This  combined sample Obs1 contains 1151
superclusters, 592 of them contain only 1 DF-cluster and were used in the
calibration of relative luminosities of superclusters (for this sample we get
$L_0 = 6.73e+11$).  The other combined observational sample Obs2 was found
using threshold density 6.0 and distance limit 520~\Mpc\ in all subsamples.

The most striking feature of the Figure is the demonstration of the presence
of numerous very luminous superclusters in observational samples, and the
absence of such systems in simulated samples.  This difference between real
and simulated supercluster richness is well seen using both richness criteria,
the multiplicity and luminosity functions. For threshold density 6.0 most
luminous simulated superclusters have a relative luminosity of about $20- 30$
in terms of the mean luminosity of richness class 1 superclusters, most
luminous superclusters of real samples have a relative luminosity about 100,
i.e. they are about 3 times more luminous.  The fraction of very luminous
superclusters (relative luminosity 20 and above) is about ten times higher in
real samples than in simulated samples.  Similar differences exist between the
multiplicity functions of real and simulated supercluster samples.  For
threshold density 6.0 the richest model superclusters have a multiplicity of
about 20 whereas real superclusters have over 50.  The number of Abell
clusters in the richest superclusters is about 30; this difference in the
number of Abell and DF-clusters can be explained by differences in the number
density of these cluster samples: the density of DF-clusters is about 2 times
higher than that of Abell clusters.  The differences between real and
simulated samples are observed not only in the region of most luminous
superclusters: over the whole richness scale the number of DF-clusters in
simulated samples is smaller than in real superclusters. 

One more interesting observation: very luminous superclusters are located in
{\em all subsamples} (Northern and Southern regions of 2dFGRS, and in
subregions of the SDSS DR4 sample, if divided into 3 wedges of equal width).
These subsamples have characterictic volumes of about 10 million cubic \Mpc,
whereas model samples of 10 times larger volume have no extremely rich
superclusters.  Kolmogorov-Smirnov test shows that the probability that real
and model distributions of supercluster luminosities and richnesses are taken
from the same parent distribution is $10^{-10}$ and $10^{-6}$, respectively.

\section{Luminous superclusters and inflation}

Superclusters of galaxies are formed by density perturbations of large scales.
These perturbations evolve very slowly.  As shown by Kofman \& Shandarin
(\cite{kofman88}), the present structure on large scales is built-in already
in the initial field of linear gravitational potential fluctuations.  Actually
they are remnants of the very early evolution and stem from the inflationary
stage of the Universe (see Kofman et al. \cite {kofman87}).  The distribution
of luminosities (or masses) of superclusters allows us to probe processes
acting at these very early phases of the evolution of the Universe.

One possible explanations for the large difference between the distribution of
luminosities of real and simulated samples is the underestimate of the role of
very large density perturbations.  If this is the case then this means that
our simulations have not yet reached a volume which can be treated as a fair
sample of the Universe. In other words, a fair sample of the Universe has
linear dimensions far in excess of 500~\Mpc, used in simulations investigated
in this Letter.  As shown by Power \& Knebe (\cite{power06}), variations in
the box size in a smaller box do not influence properties of Dark Matter
haloes in cosmological simulations.

The other feasible  explanation of the differences between
models and reality may be the presence of some unknown processes in the
very early Universe which give rise to the formation of extremely luminous
and massive superclusters.

To date it is too early to make definite conclusions on the character of
processes during inflation which may have caused the formation of very massive
superclusters.  Some aspects of this problem were recently studied by Saar et
al. (\cite{see06}).  They demonstrated that rich superclusters formed in
places where large density waves combine in similar phases to generate high
density peaks. The larger the wavelength of such phase synchronization, the
higher the richness and mass of superclusters.  The  synchronization
has properties of sound waves, where, in addition to the main frequency,
overtones appear.  In this context it is interesting to note that very rich
superclusters have a tendency to form a quasi-regular network with
characteristic scales 250 and 120 \Mpc, as demonstrated by Broadhurst et al.
(\cite{beks}) and Einasto et al.  (\cite {e1994}, \cite {e1997},
\cite{eeg97}).

The explanation of the physical origin of very massive superclusters is a
challenge for theory.  To get a more complete observational picture of the
phenomenon, large contiguous deep redshift surveys are needed. Only
contiguous  surveys allow to detect very massive superclusters.  This is one
reason why the continuation of the SDSS survey is so important, until the
whole Northern hemisphere is covered by redshift data, as originally planned.

\begin{acknowledgements}

  We are pleased to thank the 2dFGRS and SDSS Teams for the publicly
  available data releases. The Millennium Simulation used in this
  paper was carried out by the Virgo Supercomputing Consortium at the
  Computing Center of the Max-Planck Society in Garching.  The present
  study was supported by Estonian Science Foundation grants No.  4695,
  5347 and 6104 and 6106, and Estonian Ministry for Education and
  Science support by grant TO 0060058S98. This work has also been
  supported by the University of Valencia through a visiting
  professorship for Enn Saar and by the Spanish MCyT project
  AYA2003-08739-C02-01.  J.E.  thanks Astrophysikalisches Institut
  Potsdam (using DFG-grant 436 EST 17/2/05) and Uppsala Astronomical
  Observatory for hospitality where part of this study was
  performed. A.K. acknowledges funding through the Emmy Noether
  Programme by the DFG (KN 755/1). D.T. was supported by the US
  Department of Energy under contract No.  DE-AC02-76CH03000.

\end{acknowledgements}

\end{document}